\documentclass[12pt,a4paper]{article}

\usepackage{epsfig}
\usepackage{setspace}
\usepackage{subfigure}
\usepackage{graphicx}
\usepackage{longtable}
\usepackage{amsmath}
\usepackage{color}

\def\e{\begin{equation}}
\def\f{\end{equation}}
\def\%#1{\mbox{\boldmath $#1$}}
\def\=#1{\overline{\overline #1}}

\def\_#1{{\bf #1}}

\def\.{\cdot}

\def\##1{{\bf#1\mit}}

\def\am{\left(\begin{array}{c}}
\def\amm{\left(\begin{array}{cc}}
\def\a{\end{array}\right)}

\title{Microstrip antenna miniaturization using partial dielectric material filling}

\author{Olli Luukkonen, Pekka Ikonen, and Sergei Tretyakov}

\date{Radio Laboratory / SMARAD Center of Excellence\\ Helsinki University of
Technology\\P.O. Box 3000, FI-02015 TKK, Finland\\
}

\begin{document}
\maketitle {\center \large

Address for correspondence:

Olli Luukkonen, \\ Radio Laboratory, Helsinki University of Technology,\\
P.O. Box 3000, FI-02015 TKK, Finland.

Fax: +358-9-451-2152

E-mail: olli.luukkonen@tkk.fi

}

\parindent 0pt
\parskip 7pt

\vspace{0.5cm} 
\begin{center}
\section*{Abstract}
\end{center}

In this paper we study microstrip antenna miniaturization using
partial filling of the antenna volume with dielectric materials. An
analytical expression is derived for the quality factor of an
antenna loaded with a combination of two different materials. This
expression can be used to optimize the filling pattern for the
design that most efficiently retains the impedance bandwidth after
size reduction. Qualitative design rules are given, and a
miniaturization example is provided where the antenna performance is
compared for different filling patterns. Results given by the
analytical model are verified with numerical simulations and
experiments.

\textbf{Key words:} Microstrip antenna, miniaturization, partial
filling, impedance bandwidth, quality factor.

\newpage

\section{Introduction}

A microstrip antenna is nowadays one of the most commonly used antenna types. This is because of its robust
design that allows cheap manufacturing using the benefits of printed circuit board technology.

The main drawbacks of this antenna are its large size and narrow
bandwidth. Among many existing approaches to patch size reduction
(meandering the patch, introduction of shorting posts, etc.), one of
the most commonly used miniaturization methods of microstrip
antennas is loading of the antenna volume with dielectric materials
\cite{Bahl}--\cite{Balanis}. However, a dielectric loading is known
to lead to a dramatically reduced impedance bandwidth \cite{Mongia,
Hwang}.

Alternative approaches to path miniaturization may include the use of magnetic substrates, which is known to
be advantageous in terms of the antenna bandwidth \cite{hansen2000}. However, available natural
magnetic materials have rather weak magnetic properties and are rather lossy in the microwave
frequency range. The use of artificial materials (metamaterials) was recently analysed in
\cite{ikonen_ap,ikonen2005}, and it was shown that due to their dispersion the
magnetic response of the substrate does not give an advantage as compared to usual
dielectrics. As one of the possibilities, the use of non-uniform material fillings was identified in
\cite{ikonen_ap}: A non-uniform filling can modify the current distribution on the patch
and lead to increased radiation, this way compensating the negative effect of
increased reactive energy stored in the filling material. In this paper we systematically
explore this miniaturization scenario for the case of non-uniform dielectric fillings.
Earlier, partial filling was studied with the aim to reduce the antenna
size in \cite{Lee} and with the aim to modify the standing
wave pattern on the antenna element and broaden the bandwidth in \cite{Chi-Chih}.
The conclusions of these two papers are contradictory, which is another motivation for
a systematic study.
The findings of this paper are compared to the conclusions of
\cite{Lee,Chi-Chih} at the end of the paper.

First, we derive an analytical expression for the current and
voltage distribution on the antenna element loaded with a
combination of two arbitrary dispersive low-loss materials. These
expressions can be used to find a filling pattern that optimizes the
current and voltage distributions in a way that the quality factor
is minimized. We present an example miniaturization scheme where the
antenna is partially filled with a conventional dielectric material
sample located at different positions under the antenna element.
Qualitative design rules are given, and the results obtained from
the analytical model are validated by numerical simulations and
experiments.

\section{Analytical model for partially filled microstrip antennas}

In this section we derive the voltage and current distribution for a microstrip antenna loaded with a
combination of two arbitrary low-loss materials. Further, from these distributions we calculate the radiation
quality factor via the stored electromagnetic energy and the radiated power. We conduct the derivation for a
quarter-wavelength patch antenna shorted at one end, however, the model can be easily extended for half-wavelength patch antennas.

\subsection{Voltage and current distribution}

\begin{figure}[t!]
\centering \epsfig{file=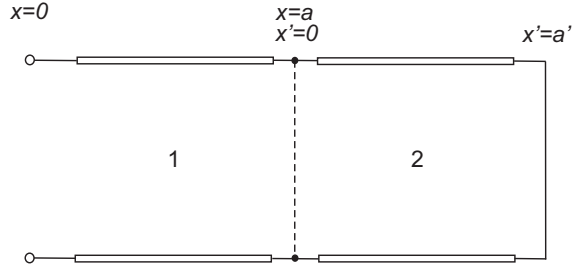,width=7.5cm} \caption{\small
Schematic illustration of a patch antenna with inhomogeneous
material filling. The substrate permittivity is different in
sections 1 and 2.} \label{fig:beginning}
\end{figure}

A microstrip antenna lying on top of a large, non-resonant ground plane and filled with two materials is
schematically illustrated in Fig.~\ref{fig:beginning}. The antenna can be modeled as a transmission-line segment
having certain characteristic impedances, shunt susceptance, and radiation conductance that depend on the
dimensions of the antenna and on the substrate materials \cite{Bahl, Balanis}. For the derivation we express
the current and voltage waves in section 1 using coordinate $x$ and in section 2 using $x'$. The coordinate
transformation is $x' = x - a$, where $a$ is the physical length of the first substrate,
Fig.~\ref{fig:beginning}.

The voltage and current distributions in both substrates can be
written as: \e U_1(x) = Ae^{-jk_1x} + Be^{jk_1x}, \f\e U_2(x') =
Ce^{-jk_2x'} + De^{jk_2x'}, \f  \e I_1(x) = Y_1\left(Ae^{-jk_1x} -
Be^{jk_1x}\right), \f \e I_2(x') = Y_2\left(Ce^{-jk_2x'} -
De^{jk_2x'}\right) ,\f
where $A$, $B$, $C$ and $D$ are the amplitudes of the waves (propagating along two directions), and $Y_1$ and
$Y_2$ are the characteristic admittances of the transmission-line segments filled with materials 1 and 2,
respectively. $k_1 = k_0\sqrt{\varepsilon_{1, \rm eff}\mu_{1, \rm eff}}$ and $k_2=k_0\sqrt{\varepsilon_{2, \rm
eff}\mu_{2, \rm eff}}$ are the wavenumbers in media 1 and 2, $\varepsilon_{\rm eff}$ and $\mu_{\rm eff}$ are
the effective substrate material parameters. The characteristic admittance can be calculated in the following
manner \cite{Collin}:
\e Z_i = \frac{1}{Y_i}= \sqrt{\frac{L}{C}} = \frac{\eta_0h}{w}\sqrt{\frac{\mu_{i,{\rm
eff}}}{\varepsilon_{i,{\rm eff}}}}, \quad i=1,2, \label{eq:impedance_zc}\f where
$\eta_0=\sqrt{\mu_0/\varepsilon_0}$ is the wave impedance of free space, $h$ is the height of the antenna
element over the ground plane, and $w$ is the width of the antenna patch.

In addition to the continuity conditions at the interface between the two substrates ($x=a$, $x'=0$), we get two
boundary conditions at the shorted edge of the patch. These conditions can be written as
\e U_1(x=a) =
U_2(x'=0),\f\e I_1(x=a) = I_2(x'=0'), \f\e U_2(a') = 0,\f\e I_2(a') = \left|I_{\rm m}\right|, \f where
$\left|I_{\rm m}\right|$ is the amplitude of the current at the shorting plate. Solving the unknown amplitude
factors $A$, $B$, $C$ and $D$ from Eqs.~(1)--(4), and using Eqs.~(6)--(9) we can write the voltage distribution
in the substrates as
\e U = \left\{
\begin{array}{ll}
j\left|I_{\rm m}\right|\left[Z_2\cos(k_1(x-a))\sin(k_2a')-Z_1\sin(k_1(x-a))\cos(k_2a')\right] & 0\leq x\leq a, \\
-jZ_2\left|I_{\rm m}\right|\sin(k_2(x'-a')) & 0\leq x'\leq a'.
\end{array} \right. \label{eq:U} \f Similarly for the current
distribution: \e I = \left\{\begin{array}{ll} Y_1\left|I_{\rm
m}\right|\left[Z_2\sin(k_1(x-a))\sin(k_2a')
+ Z_1\cos(k_1(x-a))\cos(k_2a')\right] & 0\leq x\leq a, \\
\left|I_{\rm m}\right|\cos(k_2(x'-a')) & 0\leq x'\leq a'. \end{array} \right. \label{eq:I}\f

\subsection{Stored electromagnetic energy and radiation quality factor}

At this stage we make the assumption that both substrates have low
losses. Moreover, we assume that the height of the antenna is small
and define the amplitudes of the electric and magnetic field in the
quasi-static regime as: \e E = \frac{U}{h}, \quad H = \frac{I}{w}.
\f The electromagnetic field energy density in different materials
 ($i=1$ or 2) reads \cite{jackson}:
 \e w_i^{\rm em}=\frac{\varepsilon_0\partial(\omega\varepsilon_i)}{\partial\omega}
\frac{\left|E_{{\rm m},i}\right|^2}{4} +
\frac{\mu_0\partial(\omega\mu_i)}{\partial\omega}
\frac{\left|H_{{\rm m},i}\right|^2}{4} \label{eq:elevol}, \quad
i=1,2. \f We find the electromagnetic energy stored in the
substrates by integrating Eq.~(\ref{eq:elevol}) over the antenna
volume. This leads to the following result: \e W_1 = \frac{h
w}{16k_1}\left|I_{\rm
m}\right|^2\left\{\frac{\varepsilon_0}{h^2}\frac{\partial(\omega
\varepsilon_1)}{\partial\omega}\left(\alpha_1+\alpha_2+\alpha_3\right)
+\frac{\mu_0Y_1^2}{w^2}\frac{\partial(\omega
\mu_1)}{\partial\omega}\left(\beta_1+\beta_2 +\beta_3
\right)\right\}, \label{eq:W1} \f \e W_2 = \frac{h
w}{16k_2}\left|I_{\rm
m}\right|^2\left[\frac{\varepsilon_0}{Y_2^2h^2}\frac{\partial(\omega
\varepsilon_2)}{\partial\omega}\gamma_2^- +
\frac{\mu_0}{w^2}\frac{\partial(\omega
\mu_2)}{\partial\omega}\gamma_2^+ \right] \label{eq:W2},\f where the
notations read:
\begin{center}
\begin{tabular}{l} $\gamma_1^{\pm} =  2k_1a\pm\sin(2k_1a)$, \\ \\
$\alpha_1 = Z_2^2\sin^2(k_2a')\gamma_1^+ $,\\$ \alpha_2 = -Z_1Z_2\sin(2k_2a')\left(1-\cos(2k_2a')\right)$, \\$
\alpha_3 = Z_1^2\cos^2(k_2a')\gamma_1^-$, \\ \\
$\beta_1 = Z_2^2\sin^2(k_2a')\gamma_1^- $, \\
$\beta_2 = -\alpha_2$, \\
$\beta_3 = Z_1^2\cos^2(k_2a')\gamma_1^+,$ \\ \\
$ \gamma_2^{\pm} = 2k_2a'\pm\sin(2k_2a') $.
\end{tabular}
\end{center}

The total radiation quality factor can be split into two parts: \e
Q_{\rm r} = \frac{\omega W}{P_{\rm r}} = \frac{\omega W_1}{P_{\rm
r}}+\frac{\omega W_2}{P_{\rm r}} = Q_{\rm r,1}+Q_{\rm r,2},
\label{eq:Qrtot} \f where $P_{\rm
r}=\frac{\left|U_0\right|^2}{2}G_{\rm r}$ is the radiated power,
$\left|U_0\right|$ is the amplitude of the voltage at the open edge
of the patch, and $G_{\rm r}$ is the radiation conductance. From
\cite{Balanis} we get an approximation for the radiation conductance
of a patch whose width compared to the free space wavelength
$\lambda_0$ is small: \e G_{\rm r} =
\frac{1}{90}\frac{w^2}{\lambda_0^2}. \label{eq:G} \f

Using (\ref{eq:W1}) and (\ref{eq:W2}) we get expressions for $Q_{\rm r,1}$ and $Q_{\rm r,2}$:
\e Q_{\rm r,1} =
\frac{\left|I_{\rm m}\right|^2Y_1}{8|U_0|^2G_{\rm r}}\left[\frac{1}{\varepsilon_1}
\frac{\partial(\omega\varepsilon_1)}{\partial\omega}\left(\alpha_1+\alpha_2+\alpha_3\right) +\frac{1}{\mu_1}
\frac{\partial(\omega\mu_1)}{\partial\omega}\left(\beta_1+\beta_2+\beta_3\right)\right] \label{eq:Qr1} \f \e
Q_{\rm r,2} = \frac{\left|I_{\rm m}\right|^2}{8|U_0|^2G_{\rm r}Y_2}\left[\frac{1}{\varepsilon_2}
\frac{\partial(\omega\varepsilon_2)}{\partial\omega}\gamma_2^-+\frac{1}{\mu_2}
\frac{\partial(\omega\mu_2)}{\partial\omega}\gamma_2^+\right], \label{eq:Qr2}\f $\left|I_{\rm m}\right|$ can be
expressed using Eq.~(\ref{eq:U}) and setting $x = 0$ as: \e \left|I_{\rm m}\right| = \left|U_0\right|
\left|\frac{\cos(k_1a)\sin(k_2a')}{Y_2}+\frac{\sin(k_1a)\cos(k_2a')}{Y_1} \right|^{-1}. \label{eq:Im}\f

Let us check the above formulae for the radiation quality factor by
considering a particular situation where the substrate 1 is half of
the wavelength long $\left(a=\lambda/2\right)$, and substrate 2 is
quarter of the wavelength long $\left(a'=\lambda/4\right)$. The
voltage and current distribution in substrate 1 should then
correspond to an open-ended half-wavelength patch antenna. Let us
further suppose that the patch loaded with substrate 1 would radiate
from both ends, thus, $G_{\rm r} \rightarrow2G_{\rm r}$ in
Eq.~(\ref{eq:Qr1}). $Q_{\rm r_1}$ can now be rewritten as \e Q_{\rm
r_1} = \frac{\pi Y_1}{8G_{\rm r}}\left(\frac{1}{\varepsilon_1}
\frac{\partial(\omega\varepsilon_1)}{\partial\omega}+
\frac{1}{\mu_1}
\frac{\partial(\omega\mu_1)}{\partial\omega}\right),\f which agrees
with the result derived in \cite{ikonen2005}. If we continue by
assuming that substrate 1 is dispersion-free, we get \e Q_{\rm r_1}
= \frac{\pi Y_1}{4G_{\rm r}} \label{eq:Hansen}, \f which is the
result used in \cite{hansen2000}.

\section{Impedance bandwidth behavior of partially filled microstrip antennas}

In this section we study the impedance bandwidth properties of
$\lambda/4$-patch antennas when the antenna volume is partially
loaded with different dielectric material loads located at different
positions under the antenna element. The results given by the
analytical model are verified with numerical simulations and
experiments. The known fundamental limit for the radiation quality
factor of electrically small antennas reads \cite{McLean}: \e Q_{\rm
r} = \frac{1}{k_0^3R^3} + \frac{1}{k_0R}, \label{eq:fundlimit} \f
where $k_0$ is the free-space wavenumber and $R$ is the radius of
the smallest sphere enclosing the antenna. According to
Eq.~(\ref{eq:fundlimit}), the limit does not depend on the substrate
that occupies the volume under the antenna element, because
Eq.~(\ref{eq:fundlimit}) takes into account only the fields outside
the antenna volume. This gives us the freedom to choose the
substrate and its position under the antenna element freely as long
as the material is enclosed by the sphere. Thus, for the sake of future comparisons we fix the total
volume and the resonant frequency of the antenna.
\begin{figure}[b!]
\centering \epsfig{file=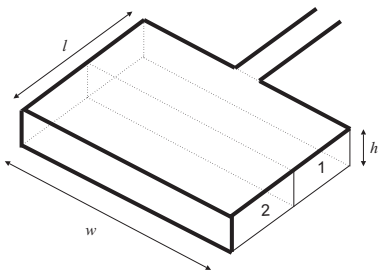,width=5cm} \caption{\small
Partially filled $\lambda/4$-patch antenna. Numbers 1 and 2 denote
different substrates.} \label{fig:positions}
\end{figure}

\subsection{Results following from the analytical model}

\begin{figure}[t!]
\centering \epsfig{file=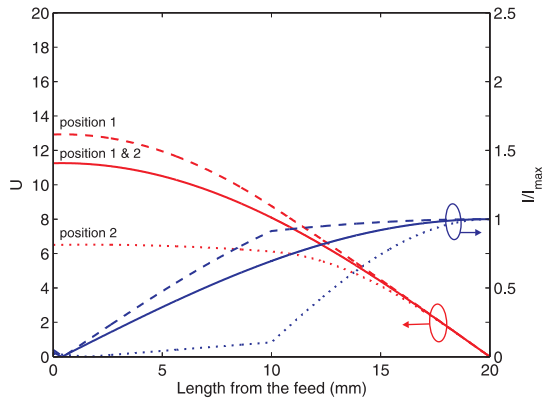,width=7.5 cm} \caption{\small The
voltage and current distributions in arbitrary units (Color figure
can be viewed in the online issue).} \label{fig:distributions}
\end{figure}

As an example, we have studied an antenna with the following dimensions: $w = 70$ mm, $l = 20$ mm, $h = 4$
mm (Fig.~\ref{fig:positions}). We consider three different filling patterns:
\begin{itemize}
\item[1)] The antenna is completely filled with a dielectric
substrate.
\item[2)] Position 1 is filled with a substrate and position 2 is empty.
\item[3)] Position 1 is empty and position 2 is filled with a substrate.
\end{itemize}
With all the filling patterns the resonant frequency of the antenna is kept at $f=2$ GHz. Thus, depending on the
filling pattern we alter the relative permittivity of the substrate as shown in Table~\ref{t}. The current and
voltage distributions corresponding to different filling patterns are shown in Fig.~\ref{fig:distributions}, and
the corresponding impedance bandwidth results (quality factors) are listed in Table~\ref{t}.

According to the results presented in Table~\ref{t}, the most optimal location for the dielectric load in terms
of the minimized radiation quality factor is position 1. It is seen in Fig.~\ref{fig:distributions} that placing
the substrate to position 1 leads to the highest voltage and current magnitudes along the patch. These higher
magnitudes increase the stored energy, however, the higher radiation voltage (open-edge voltage) increases the
radiated power. In this particular example this increase outweighs the effect of increased stored energy. When
the substrate is placed to position 2, the voltage and current magnitudes are the lowest of all the considered
cases. However, even though the amount of stored energy is the smallest, also the radiated power is small and
the radiation quality factor is the highest of all considered cases.

\begin{table}[b!]
\centering \caption{\small Impedance bandwidth results (calculated
and simulated).} \label{t}
\begin{tabular}{|c|c|c|c|}
\hline
\multicolumn{4}{|c|}{Analytical model} \\
\hline Position & $\varepsilon_{\rm r}$ & $f$ (GHz) & $Q_{0}=Q_{\rm r}$ \\
\hline 1\,\&\,2 & 3.95 & 2.00 & 29.0  \\
\hline 1    & 4.4 & 2.00 & 27.1 \\
\hline 2    & 13.7 & 2.00 & 48.1 \\
\hline \hline
\multicolumn{4}{|c|}{Simulation results (IE3D)} \\
\hline Position & $\varepsilon_{\rm r}$ & $f$ (GHz) & $Q_{0}=Q_{\rm r}$ \\
\hline 1\,\&\,2 & 3.7 & 1.99 & 27.95 \\
\hline 1    & 4.2 & 2.00 & 26.72 \\
\hline 2    & 12.5  & 2.00 & 38.79 \\
\hline
\end{tabular}
\end{table}

Results in Table~\ref{t} also show that lower permittivity
dielectrics are needed to retain the resonant frequency when the
substrate is placed near the open edge where the electric response
is the strongest.

\subsection{Simulation results}

In this section we simulate the antenna structure introduced above.
The purpose is to validate the results given by the analytical
model. We use a method of moments based simulation software IE3D.
The ground plane is infinite in the simulation setup. To ensure that
the radiation conductance is not affected by the substrate and
corresponds as closely as possible to the analytical model we leave
a 0.5\,mm long empty section before the radiating edge. The antenna
is fed with a narrow matching strip having length $l_{\rm f} = 10$\
mm and width $w_{\rm f} = 0.5$\,mm connected to a 50 $\Omega$ probe.

The fractional impedance bandwidth and the unloaded quality factor $Q_0$ can be calculated by representing the
antenna as a parallel \emph{RLC} circuit in the vicinity of the fundamental resonant frequency and using the
input voltage standing wave ratio $S$: \cite{Pues} \e BW = \frac{1}{Q_0}\sqrt{\frac{\left(T S - 1\right)\left(S
- T\right)}{S}}. \label{eq:puesBW}\f
Above, the coupling coefficient for  a parallel resonant circuit is
$T=R_0/Z_0$, where $R_0$ is the resonant resistance and $Z_0$ is the characteristic impedance of the feed line.
The voltage standing wave ratio is defined as: \e S \leq \frac{1+\left|\rho\right|}{1-\left|\rho\right|}
\label{eq:vswr}.\f We use a $\rho=-6$ dB matching criterion to define the impedance bandwidth.

The simulation results are shown in Table~\ref{t}. The simulation software gives quite accurately the
same resonant frequencies for the three different cases with nearly the same material parameters as the
analytical model.

When position~2 is filled with the substrate, the analytical model overestimates the required substrate
permittivity, and thus, predicts a higher $Q_0$. The analytical model assumes that the shorting metal plate
is perfectly conducting whereas the finite conductivity of metal is taken into account in the simulations.
When the shorting plate has a certain effective impedance, the electric response near the plate would slightly
increase due to a small increase in the electric field magnitude. This would lower the needed permittivity
value.

\subsection{Measured results}

In this subsection we present some measurement results that further
validate the analytical model. We have accurately replicated the
antenna described in the previous sections. In the measurements the
ground plane size is 30$\times$30 cm$^2$, and we load different portions of position 1 with dielectric
substrates having different permittivity values. The permittivity is
altered from case to case in order to keep the resonant frequency
fixed at 2~GHz. The loss tangent of the substrates is approximately
0.0027 in all cases.

The analytical value for $\varepsilon_1$ is calculated using Eq.~(\ref{eq:I}) and setting $x=0$. The current
amplitude at the open edge of our quarter wavelength antenna is zero at the resonant frequency. Knowing the resonant
frequency ($f_{\rm r} = 2$ GHz) and the permittivity of the material in position 2 ($\varepsilon_2 = 1$),
$\varepsilon_1$ can be solved using Eq.~(\ref{eq:I}).

\begin{figure}[t!]
\centering \epsfig{file=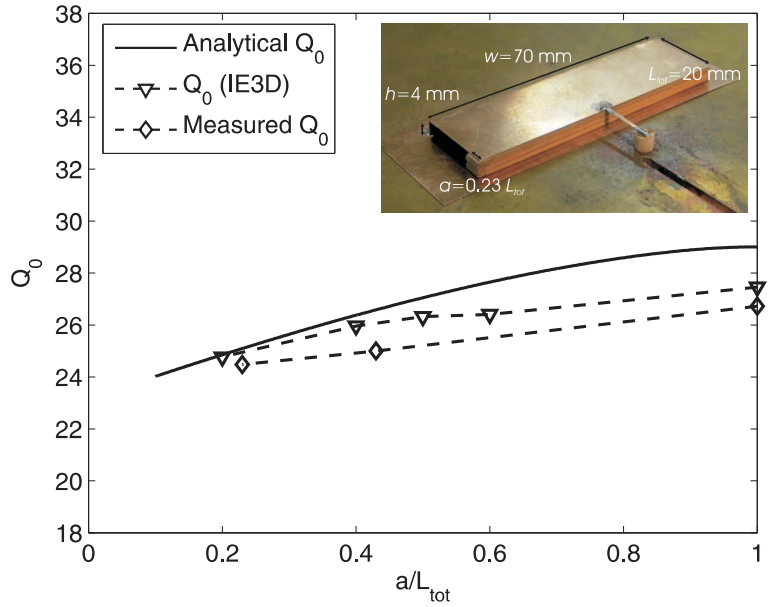, width=12cm} \caption{\small The
measured and theoretical unloaded quality factor. The photo
illustrates one of the measurement cases (Color figure can be viewed
in the online issue).} \label{fig:radQ}
\end{figure}

\begin{figure}[b!]
\centering \epsfig{file=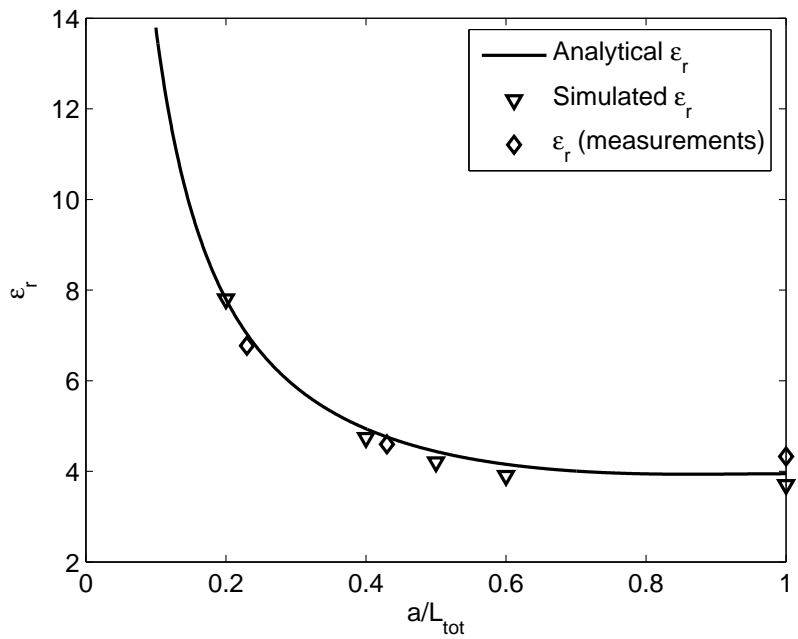, width=12cm} \caption{\small
$\varepsilon_{\rm r}$ corresponding to different filling portions.}
\label{fig:epsilon}
\end{figure}

The measured impedance bandwidths (the unloaded quality factors) are
shown in Fig.~\ref{fig:radQ} and compared with the analytical
results and simulated results. In the simulations the loss tangent
is 0.001 in all cases. The corresponding permittivity values for
each case are shown in Fig.~\ref{fig:epsilon}. The measured quality
factors and the corresponding permittivities agree well with the
simulations and with the analytical model.
The difference in the results given by the analytical model and the
measurements increase as the volume occupied by the substrate
increases. This is expected since in the analytical model the
corresponding increase in the dielectric losses is not taken into
consideration.

\subsection{Comparison with the results known from the literature}

Let us  compare the qualitative observations of our paper with the conclusions presented in
\cite{Chi-Chih}. In \cite{Chi-Chih} the authors aim to broaden the impedance bandwidth of their example antenna
by placing high permittivity dielectrics to the locations where the electric field magnitude is low, and low
permittivity dielectrics where the electric field magnitude is strong. The aim is to create a more uniform field
distribution in the antenna element. To demonstrate the feasibility of the method the authors first load a
$18\times18\times3.4$ mm$^3$ $\lambda/2$-patch antenna completely with a substrate having $\varepsilon_{\rm r}
\approx 12.6$. In the second case portions of the substrate having length $\lambda/12$ are replaced near the
open edges by a substrate having $\varepsilon_{\rm r} \approx 2.2$. The reported resonant frequencies of the
antennas are $f=2.12$ GHz and $f=2.86$ GHz. The authors compare qualitatively the impedance bandwidths and
mention that partial filling is a feasible method to efficiently miniaturize microstrip antennas.

Here we will qualitatively replicate the comparison scheme of \cite{Chi-Chih} using a $18\times9\times3.4$
mm$^3$ $\lambda/4$-patch antenna. The antenna is divided into positions 1 and 2 as in Fig.~\ref{fig:positions}.
The length of position 1 is $l/3$ and the length of position 2 is $2l/3$. When the antenna volume is filled
completely with a high-permittivity substrate having $\varepsilon_{\rm r} = 19.9$ we get from the analytical
model $f = 2.13$ GHz and $Q_{\rm r}=245$. Next we fill position 1 with a substrate having $\varepsilon_{\rm r} =
4.4$ and use the high-permittivity substrate with $\varepsilon_{\rm r} = 19.9$ in position 2. This leads to
$f_{\rm r} = 2.87$ GHz and $Q_{\rm r}=134$. We cannot, however, readily compare the quality factors as the two
antennas operate at different frequencies. When comparing the impedance bandwidth properties of two antennas
having the same volume and different resonant frequencies the proper figure-of-merit is $Q_{\rm r}\times f^3$ as
is seen from Eq.~(\ref{eq:fundlimit}). Partial filling gives a higher value for this figure-of-merit, thus, the
impedance bandwidth vs.~size characteristics of the antenna are actually better in the case of uniform filling.
If the resonant frequency of the partially filled antenna is brought to $f=2.13$ GHz we need to fill position 2
with a substrate having $\varepsilon_{\rm r} = 40$. This leads to $Q_{\rm r}=330$. Since the two antennas
operate now at the same frequency, we can directly compare the quality factors. Higher value for the partially
filled case indicates that the filling scheme proposed in \cite{Chi-Chih} is not optimal in terms of retained
impedance bandwidth. As is shown in our paper, high-permittivity dielectrics need to be positioned to the
locations where the electric field amplitude is the strongest.

The conclusion of paper \cite{Lee} is that for effective size reduction a dielectric block
should be positioned close to the radiating patch edges. This is in harmony with the conclusions from the above
analysis.

\section{Conclusions}

We have derived the voltage and current distribution for a microstrip antenna loaded with two arbitrary
dispersive and low-loss substrates. This model can be used to find the filling pattern that minimizes the
antenna quality
factor. We have presented an example miniaturization scheme where the antenna is partially filled with a
conventional dielectric material blocks located in different positions under the antenna element. Qualitative design
rules are given and the results of the analytical model are validated by numerical simulations and
experiments. It has been shown that high-permittivity dielectrics need to be positioned to the locations where
the electric field amplitude is the strongest in order to minimize the quality factor.

\section*{Acknowledgement}

The authors wish to thank Prof. Constantin Simovski and Dr. Stanislav Maslovski for their
valuable suggestions and advices.

\end{document}